\documentclass[12pt,preprint]{aastex}

\usepackage{epsfig}

\def\e10{\eta_{10}}

\def\etal{{\it et al.\ }}
\def\iso#1#2{\mbox{${}^{#2}{\rm #1}$}}
\newcommand\he[1]{\iso{He}{#1}}
\newcommand\li[1]{\iso{Li}{#1}}
\def\be#1{\iso{Be}{#1}}
\def\li#1{\iso{Li}{#1}}
\def\b1#1{\iso{B}{1#1}}

\def\beq{\begin{equation}}
\def\eeq{\end{equation}}
\def\beqar{\begin{eqnarray}}
\def\eeqar{\end{eqnarray}}
\def\simlt{\lower.5ex\hbox{$\; \buildrel < \over \sim \;$}}
\def\simgt{\lower.5ex\hbox{$\; \buildrel > \over \sim \;$}}
\def\simpropto{\lower.2ex\hbox{$\; \buildrel \propto \over \sim \;$}}

\begin{document}

\title{Higher D or Li: Probes of Physics beyond the Standard Model}

\author{Keith~A.~Olive}
\affil{ William I. Fine Theoretical Physics
Institute, School of Physics and Astronomy, \\
University of Minnesota, Minneapolis, MN 55455 USA}

\author{Patrick Petitjean and Elisabeth Vangioni}
\affil{Institut d'Astrophysique de Paris, UMR 7095 CNRS, University Pierre et Marie Curie, 98 bis Boulevard Arago,
Paris 75014, France}

\and\author{Joseph Silk}
\affil{Institut d'Astrophysique de Paris, UMR 7095 CNRS, University Pierre et Marie Curie, 98 bis Boulevard Arago,
Paris 75014, France, \\Beecroft Institute of Particle Astrophysics and Cosmology, University of Oxford, 1 Keble Road, Oxford OX1 3RH UK\\
 Department of Physics and Astronomy, 3701 San Martin Drive, The Johns Hopkins University, Baltimore MD 21218 USA }

\begin{abstract}

\vskip-6in
\begin{flushright}
UMN-TH-3037/12 \\
FTPI-MINN-12/10 \\
March 2012
\end{flushright}
\vskip+5.7in

Standard Big Bang Nucleosynthesis at the baryon density determined by
the microwave anisotropy spectrum predicts an excess of \li7 compared
to observations by a factor of 4-5.  In contrast, BBN predictions
for D/H are somewhat below (but within $~2 \sigma$) of the 
weighted mean of observationally determined values from quasar absorption systems.
Solutions to the \li7 problem which alter the nuclear processes during or subsequent to BBN, often 
lead to a significant increase in the deuterium abundance consistent with the highest values of
D/H seen in absorption systems. Furthermore, the observed 
D/H abundances show considerable dispersion.
Here, we argue that those systems
with D/H $\simeq 4 \times 10^{-5}$ may be more representative
of the primordial abundance  and as a consequence, those systems with lower D/H  would necessarily have been subject to local processes of deuterium destruction. This can be accounted for by models of cosmic chemical evolution able to destroy in situ 
Deuterium due to the fragility of this isotope. 

\clearpage

\end{abstract}


\section{Introduction}

The importance of \li7 to cosmology began with the discovery of the 
Spite plateau \citep{spitex2}. At a time when there was still considerable
uncertainty in the baryon density, \citet{yangetal} derived an upper limit
to the baryon-to-photon ratio, $\eta$, effectively using the \li7/D ratio.
Remarkably, the \li7 plateau has maintained a very constant value
over the last 30 years and over many observations, see e.g., \citep{ssm,ss2,hd,rbm,ht,thorburn,psb,ss3,mpb,bm,rnb,Asplundetal06,Bonifacioetal07,hos,hos2,sbordone,ss4}.
While the standard big bang nucleosynthesis (SBBN) predictions for 
\li7 abundances have also remained relatively stable \citep{wssok,Oliveetal00,cfo,coc02,Cocetal04,cyburt,cuoco,cfo5,iocco,coc10,coc12},
the baryon density has been determined to
unprecedented precision \citep{wmap3,wmap10} from analyses of 
microwave background anisotropies. This has led to a more 
precise prediction of the \li7 abundance that points to a clear discrepancy
\citep{cfo5,bdf} between theory and observation.

 At a baryon-to-photon ratio of $\eta = 6.16 \times 10^{-10}$ \citep{wmap10}, the BBN prediction for
 \li7/H is $(5.07^{+0.71}_{-0.62}) \times  10^{-10}$ in \cite{cfo5} and  $5.24 \times 10^{-10}$ in \cite{coc12} with an estimated error bar of 
 0.5, 
which is considerably higher than almost all observational determinations.
The value found in \citet{Ryan00} was \li7/H = $ (1.23^{+0.34}_{-0.16}) \times 10^{-10}$.
Similarly, the recent analysis of \citet{sbordone}
found \li7/H = $ (1.58 \pm 0.31) \times 10^{-10}$.
Li observations have also been performed in some globular clusters.  
For example, \citet{gonz} found  $ (2.34 \pm 0.05) \times 10^{-10}$ in NGC 6397, 
somewhat higher
than the result for field stars, whereas
\citet{monaco10} found  a value  \li7/H =$ (1.48 \pm 0.41) \times 10^{-10}$ in Omega Centauri 
 that is similar to halo star abundances.

Resolution of the \li7 problem has involved many different approaches. 
These range from questioning the nuclear reaction rates used in BBN calculations \citep{Cocetal04,angulo,cfo4,boyd}, or considering additional resonance reactions \citep{cp,chfo,brog}.
The possibility that depletion plays a role has been discussed at length \citep{vc,pinn98,pinn00,RMR05,Kornetal06,GarciaPerezetal08}.
The temperature scale used in the \li7 abundance determination has also been considered \citep{mr,hos,hos2}. There is also the possibility that the solution
of the \li7 problem requires physics beyond the standard model.  
For example, the decay of a massive particle during or after BBN could 
affect the light element abundances and potentially lower the \li7 abundance 
\citep{Jedamzik04,kkm,feng,eov,Jedamzik06,cefos,grant,cumb,kkmy,pps,jittoh,jp,ceflos,grant2,grant3,jed08a,jed08b,bjm,pp,pp2,ceflos1.5,jittoh2,kk}.
Another possibility raised recently is that of an axion condensate which cools the photon background
leading to lower value of $\eta$ at the time of BBN relative to that determined by WMAP \citep{sik,kus}. 
More exotic solutions involve the possibility of a variation in the  fundamental constants
\citep{dfw,cnouv,bfd}. 

The problem with \li7 should be put into context with respect to the other
light elements produced in BBN. In particular, there is relatively good agreement between the 
BBN predictions for \he4 and D/H and their observational determinations.
The helium abundance is the most accurately predicted of the primordial abundances. At 
$\eta = 6.16 \times 10^{-10}$, the helium mass fraction is 
$Y_p = 0.2483 \pm 0.0002$ (\cite{cfo5}); \cite{coc12} found a similar value, $Y_p = 0.2476 \pm 0.0004$.
On the observational side, the determination of the helium abundance in extragalactic HII
regions is plagued with difficulties \citep{os1}.
Using the Markov Chain-Monte Carlo methods described in \citet{aos2} and data
compiled in \citet{its}, \citet{aos3} found a higher value with
considerably larger uncertainties, $Y_p = 0.2534 \pm 0.0083$. 
Given the uncertainty, this value is consistent with the BBN prediction.

The observationally determined deuterium abundance is also in reasonable agreement with
its BBN prediction.  There are nine
 quasar absorption system observations \citep{bt98a,bt98b,omeara,pettini,lev,kirkman,omeara2,pettini2,fuma} with measurable
D/H and with a weighted mean abundance of $3.05 \pm 0.22 \times 10^{-5}$. 
This should be compared
to the BBN prediction at the WMAP value of $\eta$ 
of $2.54 \pm 0.17 \times 10^{-5}$  from \cite{cfo5} and 
$2.59 \times 10^{-5}$ from \cite{coc12} with an estimated error of 0.15.
The individual measurements
of D/H show considerable scatter (a sample variance of 0.62)
 and it is likely that systematic errors dominate the uncertainties.
While the agreement is certainly reasonable, we do draw attention to the
fact that the predicted abundance is somewhat {\em lower} than the observed mean
and this (slight) discrepancy will go in the direction of the possible \li7 solution 
discussed below.

As might be expected, when one attempts to resolve the \li7 problem by going beyond the 
standard model, there may be consequences for the other light elements.
Indeed, the destruction of \li7 is often accompanied by the production of deuterium.
In the models we will consider below, we assume a relatively late decay of a massive
particle. Specifically (to be described in more detail in the next section), we consider
the decay of a massive gravitino (mass 3-5 TeV) with a life-time of order 100-500 s.
The decays eject both hadronic and electromagnetic energy which breaks up
a small fraction of \he4 leading to an increase in the deuterium abundance and
free neutrons which destroy the freshly produced \be7 from BBN.
The models considered here were developed in \citet{ceflos}. 
In \citet{ceflos1.5}, the particular part of the parameter space
most efficient for destroying \li7 was studied and we make use of those results here.

In what follows, we will briefly review the results of post-BBN processing by
a late decaying massive particle.  We will see a tight correlation between the 
D/H  and \li7/H abundances. As a result, a
solution to the \li7 problem will invariably lead to an excess of D/H.
However, deuterium can be easily destroyed in stars.
In section 3, we first attempt to trace the evolution of these
abundances in a model of cosmic chemical evolution based on hierarchical clustering
\citep{daigne1,daigne2}.
As we will see, constraints from the cosmic star formation
rate will limit the average amount of destruction of both D and Li. 
Furthermore, we will argue that the observed scatter is probably
due to the in situ destruction of D/H. In section 4, we analyze the deuterium data 
and in section 5 we discuss new and interesting observations of HD/H2  and the 
possibility that only those absorption systems with the highest D/H represent the (post)-primordial value.

\section{Post-BBN processing of the light elements}

While there are many possible models which lead to the 
post-BBN processing of the light elements, we will focus here
on a set of supersymmetric models in which there is a massive
gravitino which decays into the lightest supersymmetric particle (LSP)
which is also a dark matter candidate \citep{Jedamzik04,kkm,kkmy,kmy,ceflos,ceflos1.5}.  
In particular, we consider a constrained version of the minimal supersymmetric model
which is described by four parameters: a universal gaugino mass, $m_{1/2}$; 
a universal scalar mass, $m_0$; a universal trilinear term, $A_0$; and the ratio of the 
two Higgs expectation values, $\tan \beta$. In addition, we must specify the sign 
of the Higgs mixing parameter, $\mu$, which take to be positive, and the gravitino mass
(see e.g. \citep{eo}).

As we will see, there is very little dependence on the choice of the specific supersymmetric model
defined by the parameters $m_{1/2}, m_0, A_0$, and $\tan \beta$. However, resulting abundances
of D and Li will be quite sensitive to the the gravitino mass and its assumed abundance. 
We will focus on a small but varied set of benchmark points \citep{bench1,bench2,bench3}
which are defined by $(m_{1/2}, m_0, \tan \beta)$ = (400,90,10) - point C; (300,1615,10) - point E; 
(460,310,50) - point L; (1840,1400,50) - point M; all with $A_0 = 0$.  The post-BBN processing of a gravitino decay for each of these benchmark points was studied extensively in \citet{ceflos,ceflos1.5} and we make use of those results here.

In the models of interest, gravitinos decay to lighter supersymmetric states, ultimately
leading to the lightest neutralino in the final state. For the post-BBN processing of 
the light elements and the destruction of \li7, decays with the injection of hadrons are most
important.  These occur through gravitino decays to a neutralino + Z, or when
kinematically allowed to a chargino + W, and gluino + gluon. Three-body decays which include
a quark-antiquark pair may also be important. Subsequent decays of the gravitino 
decay products (e.g. Z decays) lead to the injection of non-thermal protons and neutrons
which may interact with the recently produced light elements.

 The dominant primary reaction involving the non-thermal 
 decay products is the photo-erosion and spallation of \he4.
 Since \he4 dominates the abundances of the other light elements by
 many orders of magnitude, even a small amount of destruction (which leaves
 little trace on the \he4 abundance itself) may induce significant changes in the 
 abundances of the other light elements.  Generally, the spallation of \he4 leads to a
 net increase in the deuterium abundance, as one might expect. The final abundance
 of \li7 depends sensitively on the lifetime of the decaying particle. For relatively
 long lifetimes ($\ga 10^4$ s), the \li7 abundance is increased. The A=3 products of \he4 spallation, 
 tritium and \he3 undergo secondary interactions \he3($\alpha, \gamma$)\be7 and 
 T($\alpha,\gamma$)\li7, both of which increase the final \li7 abundance.
 For shorter lifetimes, there is a two-step process
 which leads to the net destruction of \li7 \citep{Jedamzik04}. First, thermalized and 
 non-thermal neutrons 
 interact with \be7 through \be7(n,p)\li7, which is followed by \li7(p, $\alpha$)\he4. 

\begin{figure}[htb]
\begin{center}
\epsfig{file=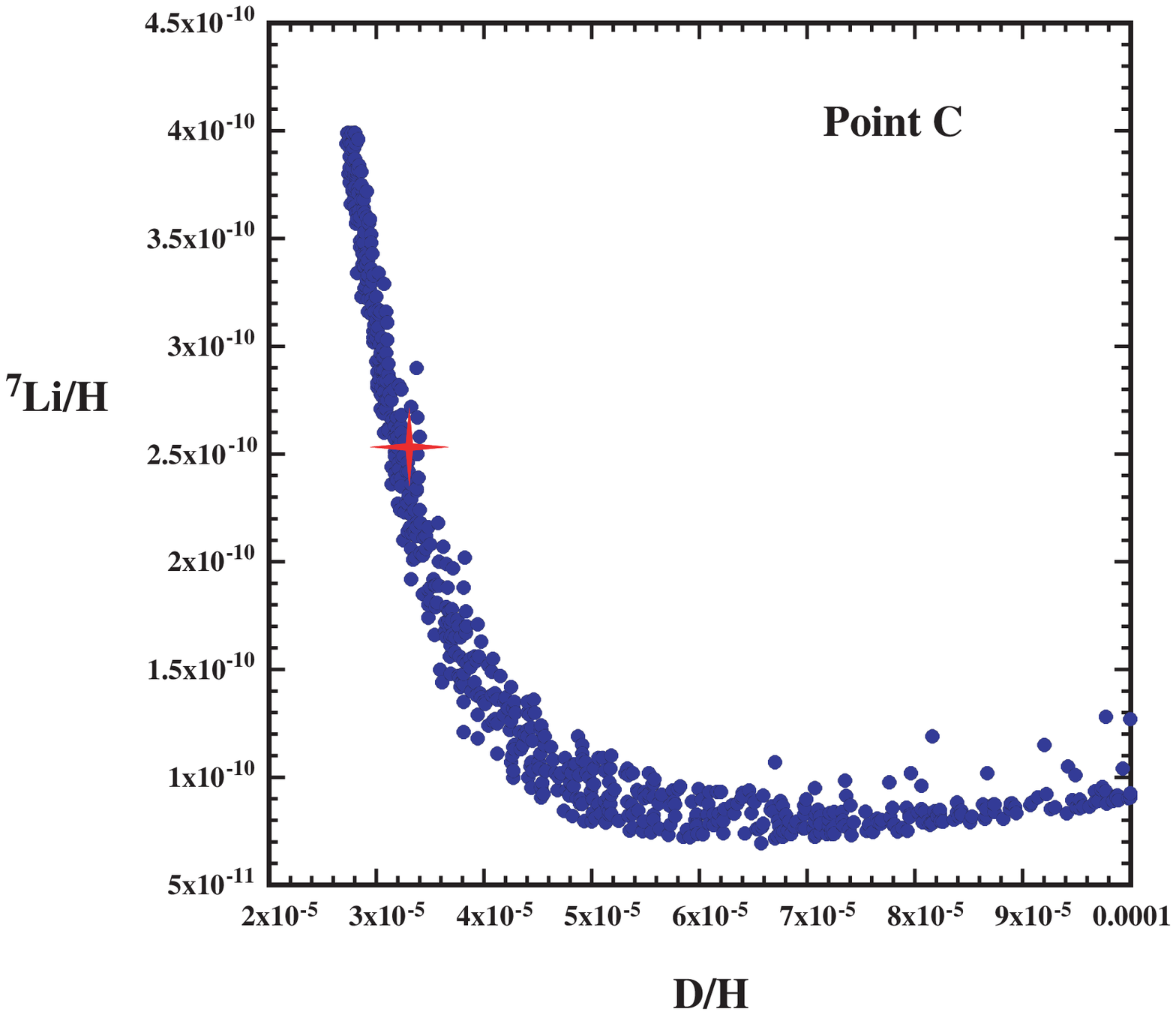, height=2.5in}
\epsfig{file=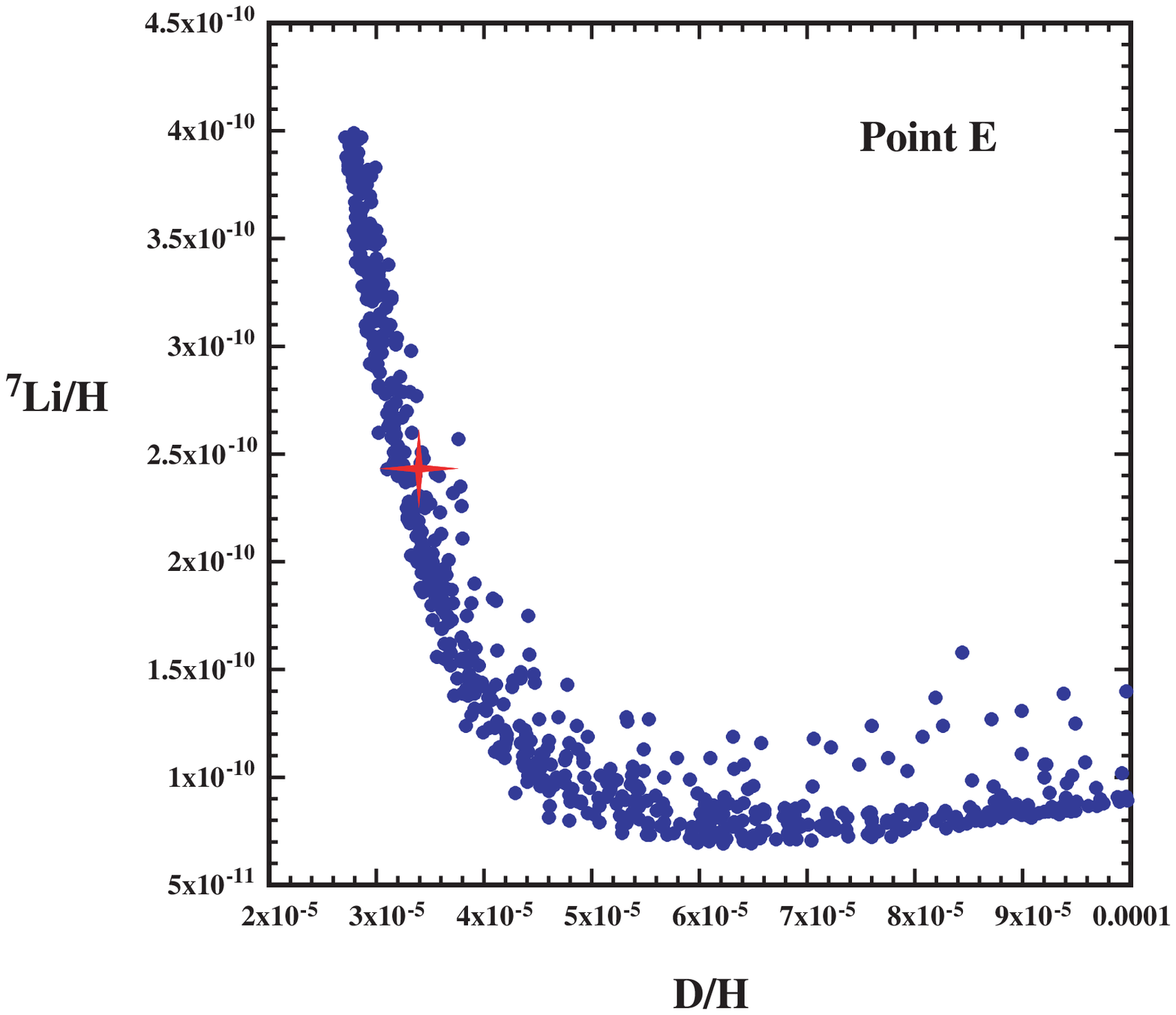, height=2.5in}
\epsfig{file=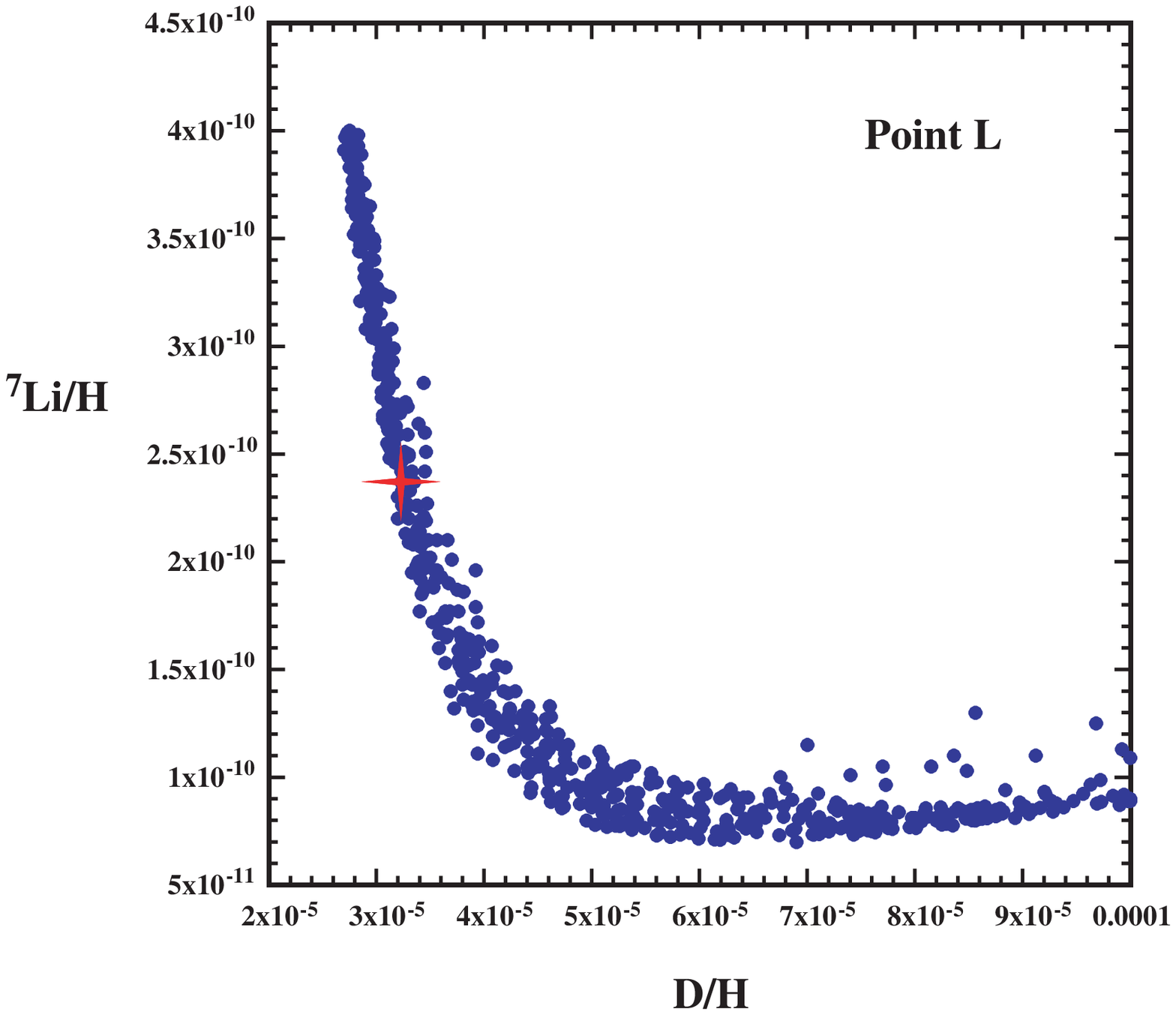, height=2.5in}
\epsfig{file=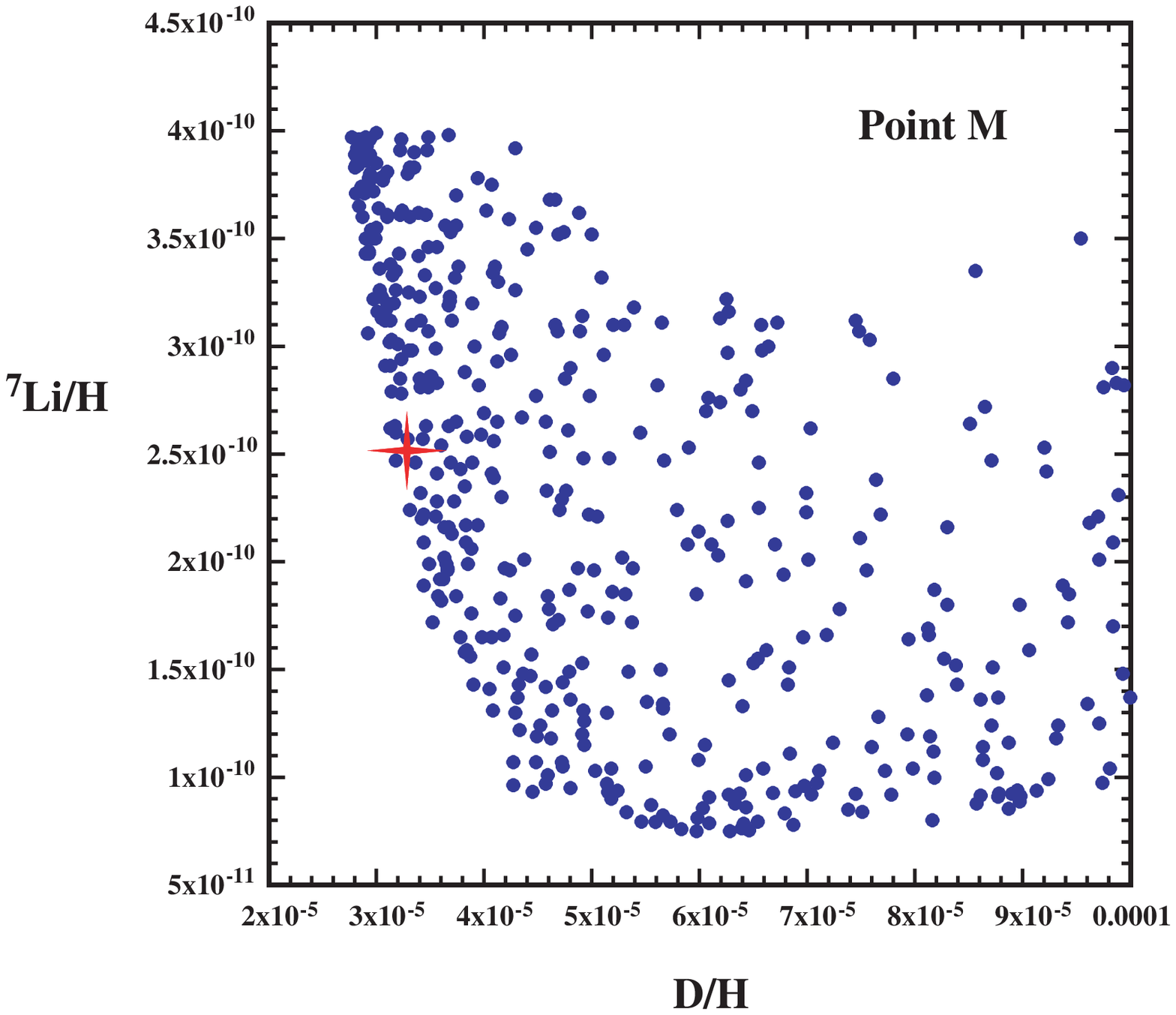, height=2.5in}
\end{center}
\caption{The resulting \li7 abundance as a function of the 
resulting D/H abundance due to the late decays of a massive gravitino.
\label{fig:dli}
}
\end{figure}

For each of the four benchmark points, \citet{ceflos1.5} used the observationally determined
abundances of \he4, D/H and \li7/H to find the value of the gravitino mass and abundance
which best fit the data. Indeed, it was found that a significant improvement to standard BBN
could be achieved with gravitino masses $m_{3/2} \sim 4 - 5$ TeV with
abundances $\zeta_{3/2} \equiv m_{3/2} n_{3/2}/n_\gamma \sim 5 \times 10^{-11} - 5 \times 10^{-10}$.
With very little dependence on the specific supersymmetric model, 
the resulting D/H and \li7/H abundances were approximately 3.2 $\times 10^{-5}$ and 
2.4 $\times 10^{-10}$ respectively. The resulting D/H and \li7 abundances for each of the 
four benchmark models is shown in Figure \ref{fig:dli} for gravitino masses in the range
2-5 TeV with abundances $\zeta_{3/2} = 2.4 \times 10^{-14} - 2.4 \times 10^{-8}$ GeV
with  cuts: \li7/H $< 4 \times 10^{-10}$ and D/H $< 10^{-4}$. The best
fit point in each case, is shown by a (red) star. While there is considerably
more scatter in benchmark point M, the lower envelope in each of the models is 
clearly very similar.

While the best fits in each case offer a significant improvement with respect to the observed abundances\footnote{The $\chi^2$ based on \he4, D/H, and \li7/H for standard BBN is 31.7.
In contrast, the $\chi^2$ for benchmark points C,E,L,M are 5.5, 5.5, 5.4, 7.0 at 
their respective best fit points \citep{ceflos1.5}.}, from Figure \ref{fig:dli}, one can see
that the \li7 abundance can be lowered further at the expense of slightly higher D/H.
However, the D/H abundance is already relatively high at the best fit points with respect to the 
weighted mean of D/H as determined from observations of quasar absorption systems as described
in the previous section.  We next examine the role of cosmic chemical evolution on these abundances
and take a critical look as to whether the weighted mean is an appropriate measure of the 
primordial deuterium abundance.

\section{The effects of cosmic chemical evolution}

\subsection{Generalities}

To follow the cosmic chemical evolution of the light elements along with
a metallicity tracer, we use an analytical model
developed first by \cite{daigne1, daigne2}, reproducing
the cosmic star formation rate in a cosmological context of structure formation.
The model is based on the standard
Press-Schechter (PS) formalism  \citep{ps} to account for non linear structures. 
The rate at which structures accrete mass is determined by
a Press-Schechter distribution function, $f_{PS}(M,z)$. 
The model tracks baryons 1) within stars or their remnants within collapsed structures,
2) in gas within collapsed structures (the interstellar medium, ISM), or 
3) outside of structures (the intergalactic medium,  IGM). 
The model includes mass (baryon) exchange between the IGM and ISM, and between 
the ISM and the stellar component.  
The age $t$ of the Universe is related to the redshift by
\begin{equation}
\frac{dt}{dz}=\frac{9.78 h^{-1}\ \mathrm{Gyr}}{(1+z)\sqrt{\Omega_\mathrm{\Lambda}+\Omega_\mathrm{m}\left(1+z\right)^{3}}}\ ,
\end{equation}
assuming the cosmological parameters of the so-called ``concordance model'', with a density of matter $\Omega_\mathrm{m}=0.27$ and a density of ``dark energy'' $\Omega_\mathrm{\Lambda}=0.73$ and taking $H_{0}=71\ \mathrm{km/s/Mpc}$ for the Hubble constant ($h=0.71$). This allows us to trace all of the quantities we describe as a function of redshift. The input stellar data (lifetimes, mass and type of remnant, metal yields, and UV flux) are taken to be dependent on both the mass and the metallicity of the star (see \cite{daigne1} for more detail).
Once these parameters are specified, we can follow many other astrophysical quantities such as the global SFR
or the abundances of individual elements (Y, D, Li, Fe, O) in the course of the expansion of the universe.
This has been used to investigate specific issues related to early star formation and the reionization epoch  \citep{rollinde}.

In this study,  we will consider a combination of three distinct modes of star formation: a normal mode of Pop II/I,  
and two additional modes of massive and intermediate mass (IM) Pop III stars (for more detail see \cite{vsof}).
Each mode has a specific IMF:
between 0.1 M$_{\odot}$ and 100 M$_{\odot}$ for the normal mode of star formation, between 2 M$_{\odot}$ and 8 M$_{\odot}$ 
for the intermediate mass mode and between 36 M$_{\odot}$ and 100 M$_{\odot}$ for the massive mode ; 
the slope of the IMF is taken to be close to the Salpeter value ($x = 1.6$).

There are several motivations for including the intermediate mass mode.
These stars may correspond to Pop III.2 stars which originate from material polluted by pristine 
PopIII.1 stars \citep{bromm2}. However, our mass range precludes any type II supernovae (SNe)
associated with this mode of star formation.  In addition, there are theoretical arguments that  the metal-free IMF predicted from opacity-limited 
fragmentation theory would peak around 4 -- 10 M$_\odot$ with steep declines
at both larger and smaller masses \citep{yoshii}.  
Primordial CMB regulated-star formation  may also lead to the
production of a population of early intermediate mass stars  at low metallicity \citep{smith09, schneider10, safra10}. 

There is considerable evidence for an early contribution by IM stars from observations. 
These stars produce very little in the way of heavy elements (oxygen and above),
but produce significant amounts of carbon and/or nitrogen and above all helium.
Evidence exists that the number of carbon-enhanced stars
increases at low iron abundances \citep{rossi} necessitating a Pop III
source of carbon, possibly in the asymptotic giant branch (AGB) phase of 
IM stars \citep{fujimoto,aoki2,lucatello} indicating possibly  an IMF peaked at
4 - 10 M$_\odot$  \citep{abia}.
In addition, the presence of s-process elements, particularly Pb
at very low metallicity also points to AGB enrichment very early on \citep{aoki,sivarani}.

 Isotopic studies of Mg also show a need for an early generation of IM stars.
While core collapse supernovae produce almost exclusively $^{24}$Mg, 
observations of  \citet{yong,yong2,alibes,Fenner03} show enhancements
(relative to predictions based on standard chemical evolution models)
in both $^{25,26}$Mg. Finally, IM stars may in part be responsible for the somewhat
high \he4 abundances seen in low metallicity dwarf galaxies \citep{vsof}. Our motivation
here for considering an early population of IM stars is clear: they offer the possibility for
destroying significant amounts of Deuterium without producing heavy elements \citep{brian}.

We fit the SFR history of Pop II/I stars to the data compiled in \citet{hb06} (from $z=0$ to 5), and to the recent 
measurements at high redshift by \citet{bouwens10} and \citet{gonzalez10}. This is modeled by
the expression suggested by \citet{sp03} :
\begin{equation}
\psi(z) = \nu\frac{a\exp(b\,(z-z_m))}{a-b+b\exp(a\,(z-z_m))}\,,
\end{equation}
where $\nu$ and $z_{\rm m}$ are the amplitude (astration rate) and the redshift of maximum SFR respectively; and
$b$ and $b-a$ are related to the slopes of the curve at low and high redshifts
respectively. The normal mode is fit using: $\nu_{\rm II/I}=0.3$ M$_\odot$ yr$^{-1}$ Mpc$^{-3}$, 
$z_{\rm m\, II/I}=2.6$, $a_{\rm II/I}=1.9$ and $b_{\rm II/I}=1.1$.
The SFR of this mode peaks at $z \approx 3$.  
These observations place strong constraints on the Pop II/I SFR.

\begin{figure}[htb]
\begin{center}
\epsfig{file=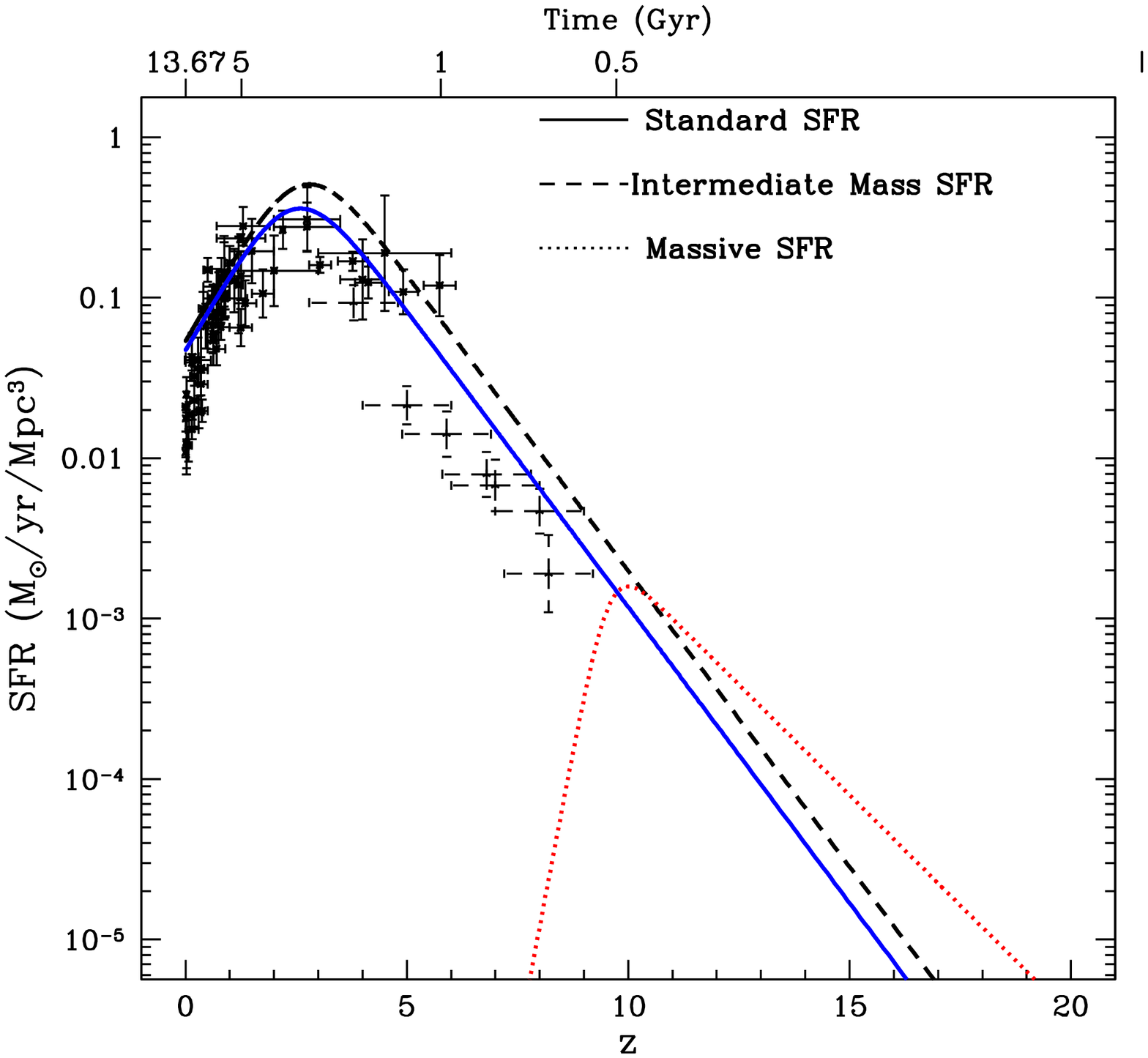, height=3in}
\epsfig{file=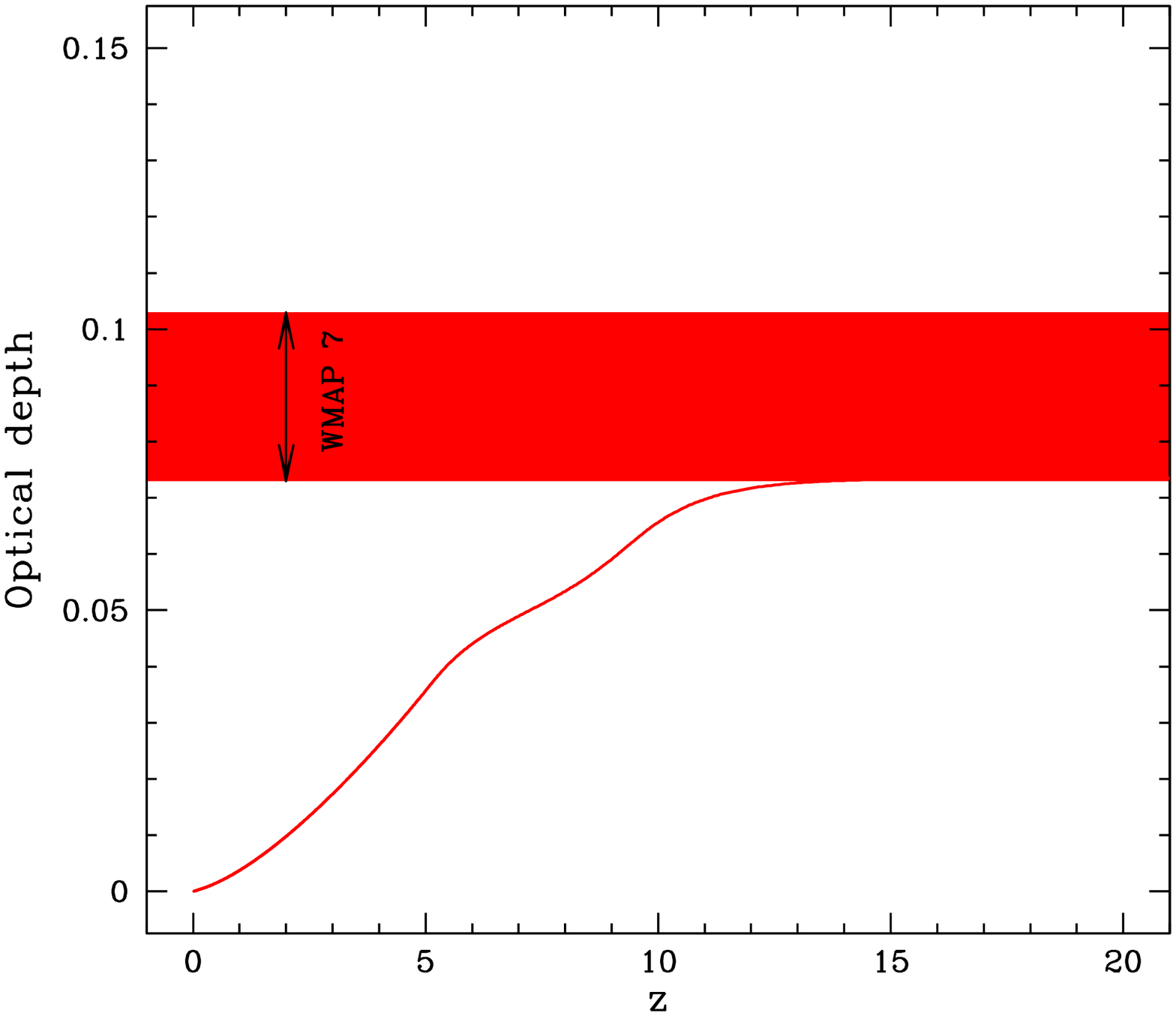, height=3in}
\end{center}
\caption{{\sl Left panel}: Cosmic star formation rate (SFR) as a function of redshift. 
The data (solid black  points and solid error bars) are taken from \cite{hb06}. Dashed black error bars 
correspond to results by \cite{bouwens10} and \cite{gonzalez10}.
The blue solid line represents the normal SFR mode with a Salpeter IMF and a mass range: 0.1 $<$ M/M$_\odot <$100.
The dotted red line represents the massive Pop III stellar mode with  a mass range: 36 $<$ M/M$_\odot<$ 100. 
The dashed black curve represents the intermediate mass SFR mode, 2$<$ M/M$_\odot<$ 8. 
{\sl Right panel}: the optical depth deduced from WMAP observations is presented as a function of the 
redshift. The red range corresponds to the observed results from WMAP7 (\cite{wmap10}). 
The red line corresponds to the model including all SFR modes. \label{fig:SFRevolution}
}
\end{figure}

As noted above, in addition to the normal mode, we add two modes for Pop III stars. 
The IM  (massive) SFR parameters are: $\nu_{\rm IIIa}=0.4 (0.7)$ M$_\odot$ yr$^{-1}$ Mpc$^{-3}$, 
$z_{\rm m\, IIIa}=2.8 (10.) $, $a_{\rm IIIa}=1.9 (4)$ and $b_{\rm IIIa}=1.1 (3.3)$.
Parameters are chosen to maximize the potential for D/H astration (without the production of 
heavy elements) while remaining concordant with the observed SFR at high redshift.
For a detailed description of the model see \citet{rollinde} and \citet{daigne2}. A recent study
shows that the first stars have feedback-limited masses of around 40 M$_\odot$
\citep{hosokawa2011}, appropriate for core-collapse supernovae.

In Figure \ref{fig:SFRevolution}, we show the adopted SFR for each of the three modes 
considered. As one can see, the normal mode and the intermediate mass SFR
 are constrained to fit the observations plotted in the figure. As the data extend only up 
to $z \approx 8$, there is  little constraint for $z > 8$. Indeed, data at these redshifts  are highly uncertain due to unknown systematics involving, among other effects, dust correction and adopted rest-frame UV luminosity function
\citep{labbe10}. In this context, the massive mode which can dominate at $z \ga 10$ is constrained to fit the optical depth deduced from WMAP observations (shown in the right panel of Figure \ref{fig:SFRevolution} ).The global SFR used in this study is the sum of these three modes.

\subsection{The cosmic evolution of D and Li}

In the cosmological context described above, it is possible to track the evolution of both light elements, D and \li7. 
Following the analysis of Section~2 corresponding to the post-BBN processing of the light elements, we consider 
the resulting D and \li7  abundances due to the late decays of a massive gravitino (see Figure 1). 
As noted earlier, a general consequence of lowering \li7 is  higher D/H. So to fit the Spite plateau we choose three 
representative values for the \li7  abundance corresponding to the values quoted in the introduction:  \li7/H = $1.23 \times 10^{-10}$, $1.58 \times 10^{-10}$, and  $2.34 \times 10^{-10}$. 
These abundances 
correspond to D/H = $4.4 \times 10^{-5}$, $3.9\times 10^{-5}$, and $3.3\times 10^{-5}$ respectively.

\begin{figure}[htb]
\begin{center}
\epsfig{file=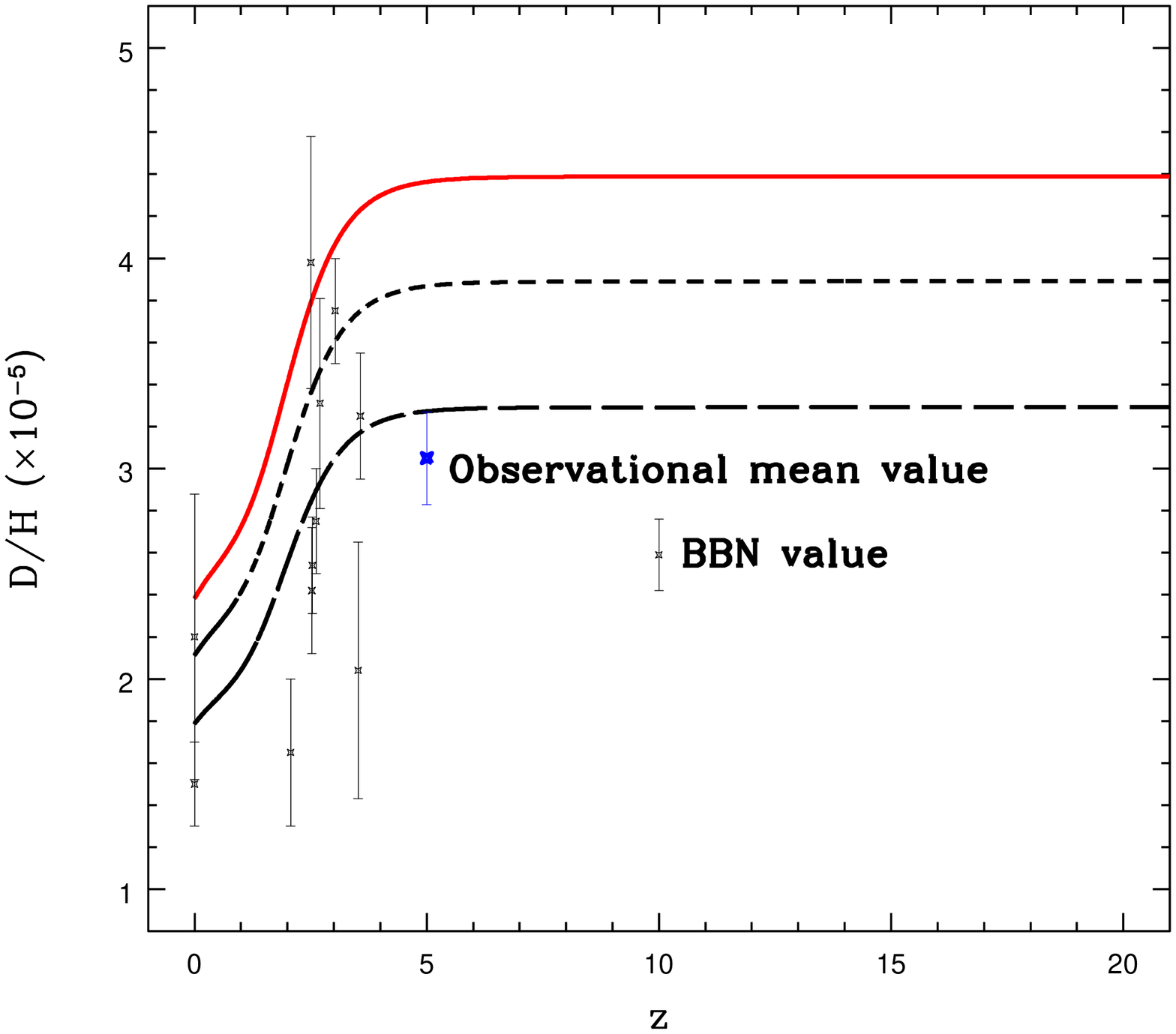, height=3in}
\epsfig{file=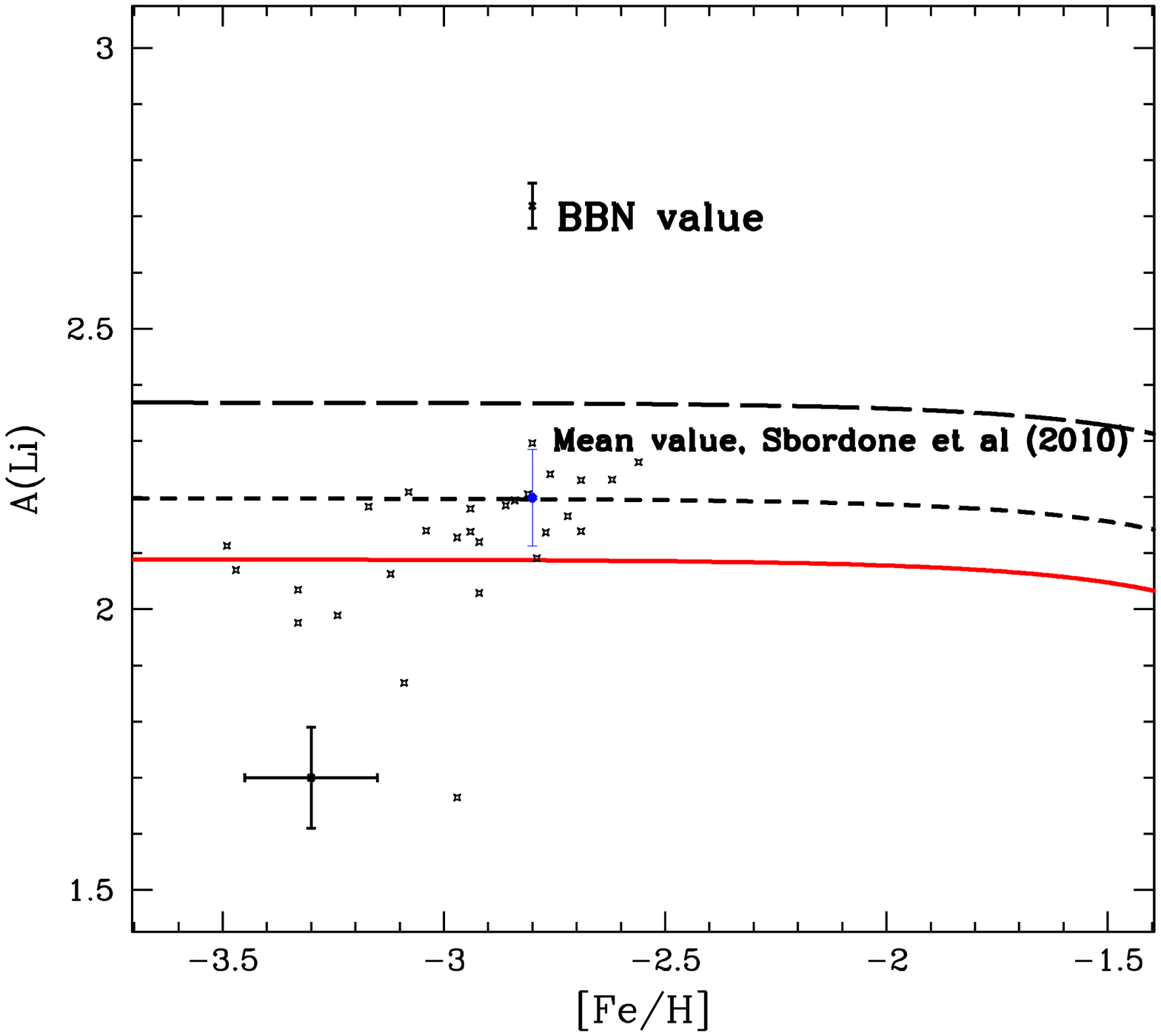, height=3in}
\end{center}
\caption{D/H abundance as a function of redshift (left) and A(Li)~=~log$(Li/H) + 12$ as a function of [Fe/H] (right). 
Deuterium data come from observations described in the text.  Lithium data come from \citep{sbordone}. 
The red solid (black short dashed and large dashed) lines correspond to initial post-BBN Li/H, D/H values:  $1.23 \times 10^{-10}$, $4.4 \times 10^{-5}$,  ($ 1.58 \times 10^{-10}$ , $3.9\times 10^{-5}$, 
and $2.34 \times 10^{-10}$, $3.3\times 10^{-5}$) respectively.
\label{fig:dlievolution}
}
\end{figure}
 
In Figure \ref{fig:dlievolution} we show the cosmic evolution of lithium  as a function of metallicity and the 
corresponding D/H evolution as a function of redshift  in the three selected cases. 
At the metallicities of  interest, there is very little evolution in the \li7 abundance.  
The high redshift Deuterium data come from 
observations discussed in the next section. 
The local Deuterium data ($z = 0$) is taken from \citet{lin} in the Local Bubble
and the higher value from \citet{sav} in the warm neutral medium of the Galactic halo.
D destruction begins at redshift 3, corresponding to the peak of the 
cosmic SFR (see figure 2, left). The global evolution of D/H fits the observations reasonably well, given the observational
uncertainties (see next Section for a discussion).

It is important to note the high dispersion of the observed values at $z\sim 3$
(corresponding to $t$~=~3$-$4~Gyr) in the range  $ 2.4 - 4. \times 10^{-5}$. 
This could be a consequence of different star formation histories in the galaxies associated with the DLAs. Note that the individual chemical evolution histories are not probed by
our model of cosmic chemical evolution which only yields the average abundances of
the elements in structures and in the IGM. However, 
D is a very fragile isotope (it is destroyed in stars at $T$~=~$10^5$~K) and
its destruction rate 
is highly dependent on the ratio $\sigma$ = gas mass/total mass of the galaxy 
as shown in \citet{vangioni88} (see their figure 3); a low ratio corresponding to a low D abundance. As a consequence, D/H destruction factors can range between 2 to  10.  Indeed,  \citet{brian} showed
that D can be efficiently destroyed without overproducing metals 
in a standard chemical evolutionary model 
 with an early population of intermediate mass stars (see their figure 1). 
It is reasonable to think that these pristine structures are in different evolution stages 
and consequently have different gas masses and processing histories. Finally,  as D can only be destroyed in the course of chemical evolution \citep{els}, it is reasonable to relate the highest D/H observation to 
the primordial post-BBN abundance.
In this context, we are able to reconcile the BBN D abundance and  high D/H observational values at high 
redshift together with  theoretical and observational Li abundances. 
 
\section{Analysis of the D/H observations}

Most of the measurements available in the literature have been gathered
by \citet{pettini2} and more recently in \citet{fuma}. It was
concluded, however, that the measurements by \citet{lev}
and \citet{cri} are not relevant because not all of the D~{\textsc i}
components in those observations are resolved.
This effect would imply systematically low D/H if one 
has included velocity components in H~{\textsc i} that are not detected
in D~{\textsc i} because the D~{\textsc i} is hidden by other H~{\textsc i} absorption features
\citep{kirkman}.

 \citet{lev} convincingly showed that D~{\sc i} is detected at
$z=3.03$ toward Q~0347$-$3819 in the transitions Ly-8, 10 and 12 (see their
Fig.~12). They fit the D~{\sc i} absorption with a profile that is
consistent with the N~{\sc i} and O~{\sc i} profiles.
Although the latter is saturated, the former is well defined
in N~{\sc i}$\lambda$$\lambda$953.4,964.0 (see their Fig.~10).
The high D~{\sc i}/H~{\sc i} value derives from the presence of a narrow
component with Doppler parameter as small as $\sim$3~km~s$^{-1}$. The
presence of this narrow component is well demonstrated however by the
detection of H$_2$ in a single component. Thirty one H$_2$ transitions are
used to derive the Doppler parameter from the curve of growth (see their
Fig.~3). In addition the H$_2$ excitation diagram indicates a temperature
of $\sim$825~K which is consistent, within errors, with the low value of
the Doppler parameter. The temperature found here is also consistent with
the findings that the molecular fraction is small in most DLAs because the
gas is warm \citep{petitjean2006}.
Although errors may have been
underestimated (see Appendix of \citet{kirkman},
 there is no objective
reason to reject this measurement, D/H~=(3.75$\times$0.25)$\times$10$^{-5}$
for log~$N$(H~{\sc i})~=~20.56$\pm$0.05 and Zn/H~=~0.98$\pm$0.09 \citep{led}.
Thus,  we have included this value in our study. Note that 
\citet{dodo} report a lower value in this system: D/H~=~2.24$\pm$0.67$\times$10$^{-5}$,
but their $b$ values are certainly too large and their value should be considered as
a lower limit.

\citet{cri} reported D/H~=~1.6$^{+0.25}_{-0.3}$$\times$10$^{-5}$ 
towards PKS~1937$-$1009 in a Lyman Limit System (LLS) 
with log~$N$(H~{\textsc i})~=~18.25$\pm$0.02 and 
for a metallicity of [Si/H]~=~$-2.0\pm 0.5$ at $z= 3.256$.
In that case only Lyman-$\alpha$ and Lyman-$\beta$ lines are used and indeed,
a model with lower H~{\textsc i} and higher D~{\textsc i} column densities can be
accommodated (see \citet{omeara2}). As a result, we have not included this reported 
measurement in our analysis.

In addition, it is surprising that \citet{pettini2} do not consider
the measurement by \citet{lev} when they retain the measurement at $z=2.06$ towards
Q~2206$-$199 which is obtained from a low SNR intermediate resolution
($R\sim 15000$) STIS-HST spectrum (see discussion in the
Appendix of \citet{kirkman}).  
Furthermore, the $b$ value and the decomposition
of the profile are uncertain in this observation and the D/H ratio has been claimed to be possibly as
high as 2.9$\times$10$^{-5}$ (see  \citep{kirkman}). While we retain this object in our average,
we comment as well on the effect of excluding it.

Very recently, \citet{fuma} measure log~(D/H)~=~$-$4.69$\pm$0.13 in 
a LLS (log~$N$(H~{\textsc i})~=~17.95) at $z=3.522$ towards SDSS J113418.96+574204.6.
They argue that the gas is pristine with metallicity $<$10$^{-4.2}$ solar.
However, at such a low neutral hydrogen column density, the gas is most probably 
partially ionized and a correction for ionization effects has to be applied. 
To reach their conclusion, they assume a value for 
the ionization parameter (log~$U$~$>$~$-3$) and argue that a gas density
larger than $n_{\textsc H}$~$>$~10$^{-2}$~cm$^{-3}$ is unusual in LLSs. We believe that this 
assumption is unjustified and a density of $n_{\textsc H}$~$\sim$~10$^{-1}$~cm$^{-3}$
is perfectly acceptable (see e.g. \citet{pbp}). Therefore 
a robust upper limit on the metallicity is not less than 10$^{-3}$ solar. This abundance
is typical of what is found for Damped Lyman $\alpha$ Systems (DLAs) 
with the smallest observed metallicities 
\citep{penprase,cooke}. Therefore, the gas may not be pristine and the low
value of the D/H ratio could be a consequence of deuterium destruction by star formation
activity.
 
Consequently, we have included all deuterium measurements from \citet{pettini2} plus
the measurements by \citet{lev} and the recent report in \citet{fuma}.
The weighted mean of these nine measurements is D/H = $(3.05 \pm 0.22) \times 10^{-5}$,
where a scale factor of $S = 2.0$ ($= \sqrt{\chi^2/8}$) is applied to the the error in the mean 
to partially account for the large dispersion.
The sample variance of would imply a much larger uncertainty of 0.62 $\times 10^{-5}$.
We note that this value is in modest {\em disagreement} with the SBBN value 
of $2.54 \pm 0.17 \times 10^{-5}$  from \cite{cfo5} or
$2.59 \pm 0.15 \times 10^{-5}$ from \cite{coc12}.
Excluding the data from \citet{lev}, we find D/H = $(2.80 \pm 0.20) \times 10^{-5}$ ($S = 1.64$)
and a sample variance of (0.52 $\times 10^{-5}$) for eight objects. Neglecting instead the data
from \citet{pettini}, we find  D/H = $(3.11 \pm 0.21) \times 10^{-5}$ ($S = 1.83$)
and a sample variance of (0.55 $\times 10^{-5}$), also for eight objects. 

It has already been noted by several authors that the dispersion in
the reported D/H-values is much larger than what is expected from individual errors.
This could be a consequence of the errors having been underestimated.
The presumption of a unique value for D/H, however, is
not supported by the observations (see also \citet{ivan}).
Furthermore, if the post-BBN value for D/H were at or near the upper end of the existing data 
as in Figure \ref{fig:dlievolution}, the dispersion seen in the data
could be explained by the in situ destruction of D/H through chemical 
evolution. If this is indeed a local effect, the extent of the dispersion is not surprising.

\section{Discussion}

As mentioned in the Introduction, there are many potential solutions to the 
SBBN \li7 problem. It is fair to say that all of them require some additional
input beyond what we normal term as standard.  This may involve additional
turbulent features in diffusive models of stellar evolution \citep{Kornetal06} or
new resonant reactions  \citep{cp,chfo,brog}.  However, several of the possible solution
come with the price of an increased D/H abundance: photon cooling \citep{sik,kus},
variable fundamental constants \citep{dfw,cnouv,bfd} and particle decays during or 
after BBN 
\citep{Jedamzik04,kkm,feng,eov,Jedamzik06,cefos,grant,cumb,kkmy,pps,jittoh,jp,ceflos,grant2,grant3,jed08a,jed08b,bjm,pp,pp2,ceflos1.5,jittoh2,kk} all tend to increase D/H while yielding a 
\li7 abundance which matches the low metallicity Pop II abundance determinations.
Here, we have used the results of \citet{ceflos1.5} for the specific correlations
between the post-BBN D/H and \li7/H abundances as displayed in Figure \ref{fig:dli}. 

Since the D/H abundance is expected to decrease monotonically over time,
a high value of D/H from BBN is not necessarily problematic.
First of all, we emphasize that SBBN predicts a value of D/H which is about 2$\sigma$ 
below the weighted mean of observational determinations. Second, while the post-BBN 
abundance of D/H may exceed the observational value, we expect
some destruction of deuterium due to chemical evolution. 
Indeed in a model based on hierarchical clustering, we have seen that there is some
modest astration of D/H while leaving \li7 virtually unperturbed as seen in Figure \ref{fig:dlievolution}. 
However, because these models yield average abundances, they can not account for the 
observed dispersion seen in the data. As noted above, this may be due to either
under-estimated errors or the in situ destruction of D/H. 

Before concluding, we note that
the HD/H2 ratio is an interesting alternative for D/H investigations in DLAs. 
In the diffuse ISM, with physical conditions very similar to those of DLAs, 
the formation of HD occurs via the reaction:
H$_2$~+~D$^+$~$\to$~HD~+~H$^+$ while its destruction is due to photodissociation.
Because of the low abundance of deuterium, the transition between atomic deuterium and HD 
takes place deeper in a cloud than the transition between atomic and molecular hydrogen. 
Moreover, while the H$_2$ column density is usually very well constrained
by the presence of numerous transitions, only a few transitions are seen for HD,
making these measurements quite uncertain. 
Therefore, although the ratio should be interpreted with care (e.g. \citet{tum}),
$N$(HD)/2$N$(H$_2$) in the cloud should give a lower limit on the D/H ratio (see \citet{lacour}). 

Molecular hydrogen is found in less than 10\% of DLAs \citep{led,notrea}
and to date, only five detections of HD at high redshift have been reported.
Interestingly, some high D/H values are observed despite significant metal enrichment:
$N$(HD)/2$N$(H2) = 1.5$\times$10$^{-5}$, 3.6$\times$10$^{-5}$, 7.9$\times$10$^{-5}$,
1.6$\times$10$^{-5}$ and 0.95$\times$10$^{-5}$ at $z_{\rm abs}$ = 
2.42, 2.33, 2.10, 2.63, and 2.69 
towards, J~1439+1117 \citep{notrea},
Q~1232+082 \citep{ivan}, J~2123$-$0500, FJ~0812+32
\citep{tum} and SDSSJ 123714.60+064759.5 \citep{notreb} for metallicities
relative to solar [X/H]~=~+0.16, $-$1.43, +0.36, $-$0.48 and +0.34 respectively.


It is apparent that these measurements, although uncertain, are all consistent
with the cosmological D/H ratio and, more importantly, a factor of ten above 
what is measured in our Galaxy \citep{lacour}.
This finding is somewhat puzzling, in light of the diversity in the other properties of these
systems and in particular their high metallicity. 
Infall of primordial gas can help maintain such high D/H ratios in the course
of galactic evolution (e.g., \citet{pf}). However, such process may have
difficulties alone to explain the observations. Consequently, these $HD/H{_2}$ observations seem
 to indicate that D/H values at hight redshift are somewhat high; indicating again that a high primordial deuterium value  is privileged.

In conclusion, we have argued that although many models which
are capable of reducing the high BBN abundance of \li7 to the level seen in Pop II halo stars
come with a price of an increase in the D/H abundance, this may well be a feature and not a bug.
The weighed  mean of the D/H abundance data (modestly) exceeds the SBBN value,
and carries considerable dispersion. If the upper envelope of this data reflects the post-BBN 
abundance of D/H, then these models which attempt to resolve the \li7 problem are preferred.
The dispersion seen (if real) must then be explained by the astration of D/H in local
processes.  We note in closing that non-standard model explanations of the primordial \li7 abundance generically predict \li6 in excess of the standard model prediction as well as an increase in the deuterium abundance. Confirmation of  \li6  in extremely metal-poor halo stars would provide significant support for models in which nuclear processes during or subsequent to BBN resolve the \li7 problem.

\label{end}

\acknowledgements
We would like to thank Feng Luo for help with the data leading to Figure 1. 
The work of KAO was supported in part by DOE grant
DE--FG02--94ER--40823 at the University of Minnesota. 
This work was also supported by the PICS CNRS/USA and the french ANR VACOUL.

\end{document}